\documentclass[prb,superscriptaddress,preprint,showpacs]{revtex4}
\usepackage{graphicx}
\usepackage{dcolumn}
\usepackage{bm}

\newcommand{\peakkkk}{\ensuremath{\langle k k k\rangle}}
\newcommand{\peakkk}{\ensuremath{\langle k k 0\rangle}}
\newcommand{\peakk}{\ensuremath{\langle k 0 0 \rangle}}
\newcommand{\sigmapi}{\ensuremath{\sigma\rightarrow\pi}}
\newcommand{\sigmasigma}{\ensuremath{\sigma\rightarrow\sigma}}
\renewcommand{\vec}{\bm}

\begin{document}

\preprint{SBW/KKK/New version 2}

\title{Multi-$\vec{k}$ configurations}

\author{N. Bernhoeft}
\affiliation{\'{e}partment de Recherche Fondamentale sur la
  Mati\`{e}re Condens\'{e}e, CEA, F-38054 Grenoble CEDEX, France\\}
\author{J. A. Paix\~{a}o}
\affiliation{Physics Department, University of Coimbra,
  Coimbra, 3004--516 Portugal\\}
\author{C. Detlefs}
\affiliation{European Synchrotron Radiation Facility,
  Bo\^\i te Postal 220, F-38043 Grenoble CEDEX, France\\}
\author{S. B. Wilkins}
\affiliation{European Commission, JRC, Institute for
  Transuranium Elements, Postfach 2340, Karlsruhe, D-76125 Germany\\}
\affiliation{European Synchrotron Radiation Facility,
  Bo\^\i te Postal 220, F-38043 Grenoble CEDEX, France\\}
\author{P. Javorsky}
\affiliation{European Commission, JRC, Institute for
  Transuranium Elements, Postfach 2340, Karlsruhe, D-76125 Germany\\}
\author{E. Blackburn}
\affiliation{European Commission, JRC, Institute for
  Transuranium Elements, Postfach 2340, Karlsruhe, D-76125 Germany\\}
\affiliation{Institut Laue-Langevin, Bo\^\i te Postal 156,
  F-38042 Grenoble CEDEX, France\\}
\author{G. H. Lander}
\affiliation{European Commission, JRC, Institute for
  Transuranium Elements, Postfach 2340, Karlsruhe, D-76125 Germany\\}
\date{\today}

\begin{abstract}
Using resonant x-ray scattering to perform diffraction experiments 
at the U M$_{4}$ edge novel reflections of the generic form 
\peakkkk\ have been observed in UAs$_{0.8}$Se$_{0.2}$ where $\vec{k} 
= \peakk$, with $k = \frac{1}{2}$ reciprocal lattice units, is 
the wave vector of the primary (magnetic) order parameter. The 
\peakkkk\ reflections, with $10^{-4}$ of the \peakk\ magnetic 
intensities, cannot be explained on the basis of the primary 
order parameter within standard scattering theory. A full 
experimental characterisation of these reflections is presented including their
energy, azimuthal and temperature dependencies. On this basis 
we establish that the reflections most likely arise from the electric dipole 
operator involving transitions between the core $3d$ and partially 
filled $5f$ states. The temperature dependence couples 
the \peakkkk\ peak to the triple-$\vec{k}$ region of the phase 
diagram: Below $\sim 50$~K, where previous studies have suggested 
a transition to a double-$\vec{k}$ state, the intensity of the \peakkkk\ is
dramatically reduced. Whilst we are unable to give a definite  explanation of
how these novel
reflections appear, this paper concludes with a discussion of possible ideas
for these 
reflections in terms of the coherent
superposition of the 3 primary (magnetic) order parameters. 
\end{abstract}
\pacs{75.25.+z,75.10.-b}
\maketitle

\section{INTRODUCTION}

Since their discovery in 1963 by \citeauthor{Kouvel63} \cite{Kouvel63}, multi--$\vec{k}$
configurations have generated their share of confusion in the 
description of magnetic structures. The ambiguities arise since 
magnetic systems commonly lower their free energy by formation 
of domains. This eventuality frequently renders the best-known 
technique for their microscopic identification, neutron diffraction, 
impotent in the determination as to whether the magnetic structure 
is single-$\vec{k}$ or multi-$\vec{k}$. The respective characteristics are
that a single-$\vec{k}$ 
configuration has only one magnetic propagation vector in any given 
magnetic domain whilst a multi-$\vec{k}$ configuration is defined by the 
\textit{simultaneous} presence of more than one such propagation vector. 
Given the possibility of multiple 
scattering, one immediately sees the likelihood of confusion 
in the interpretation of diffraction peak intensities from a 
multi-domain sample.

Neutron diffraction is a bulk technique, sensitive to 
the spatial periodicities of the magnetic field modulation. 
In general one cannot locate the scattering 
volume from which the diffraction peaks arise to a precision 
better than that given by the incident and scattered beams' intersection 
with the sample. Given incident flux limitations, even at the 
most powerful neutron sources, beams can rarely be made sufficiently 
small (on the scale of magnetic domains) to identify unambiguously
the magnetic configuration from intensity measurements in a multi-domain 
sample. External perturbations can, of course, change the domain 
populations and may allow identification, but this always begs
the question as to whether the external perturbation may have 
changed the intrinsic magnetic configuration.
\begin{figure}[p!]\begin{center}
\includegraphics[height=0.65\textheight]{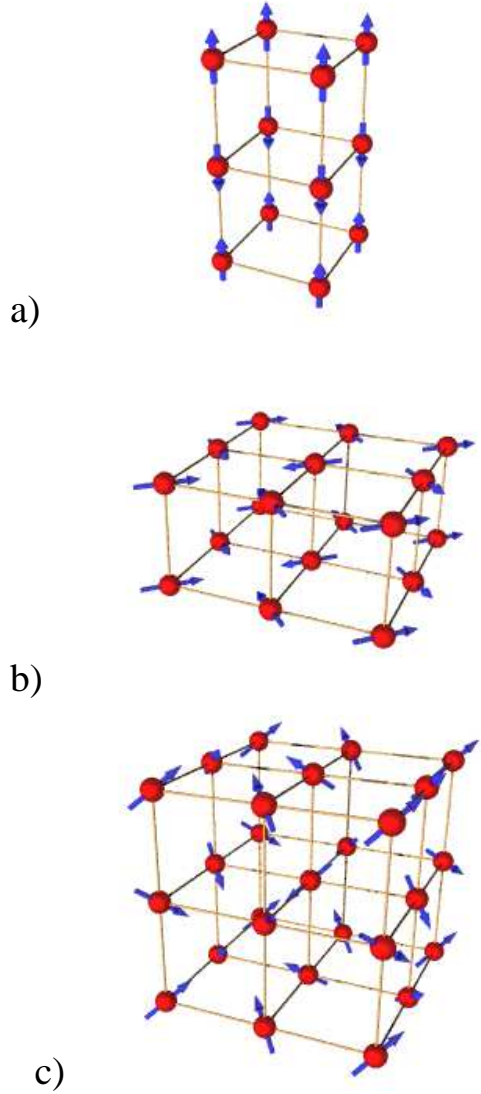}
\label{fig:cd-povray}
\caption{UAs$_{1-x}$Se$_x$ crystallizes in the fcc NaCl structure. The
  magnetic modulation wavevector is \peakk\ where $\langle\ldots\rangle$
  signifies all permutations of k, and in the composition discussed here $k =
  \frac{1}{2}$.  The magnetic moments of the four uranium atoms within the
  chemical unit cell have the same direction and magnitude. For simplicity we
  have therefore 
  shown only the magnetic moment in the corner of each chemical unit cell. (a)
  In a longitudinal single-$\vec{k}$ structure with $\vec{k} = [0 0 1/2]$, the
  moments are aligned along $[001]$ and the unit cell is doubled along this
  direction. (b) In the $2\vec{k}$ structure with $k_a = [100]$ and $k_b =
  [010]$, the net moment is along $[110]$, and the unit cell is doubled in the
  $a$ and $b$ directions. (c) In the $3\vec{k}$ structure the unit cell is
  doubled in all directions and the net moment direction is $[111]$. Of all
  the structures the $3\vec{k}$ is the only one in which a unique domain
  exists.}
\end{center}\end{figure}

To be specific we take the case of the system UAs$_{1-x}$Se$_{x}$ where 
complete solid solutions exist and a considerable amount of neutron 
diffraction has been performed \cite{Kuznietz87}. Diagrams of the possible 
magnetic structures for $0 < x < 0.3$ are shown in Figure~\ref{fig:cd-povray}. The single-$\vec{k}$ configuration has 3 distinct (tetragonal) domains, 
as does the $2\vec{k}$ state, whilst the cubic $3\vec{k}$ phase forms 
in a single domain. In these illustrations the repeat distance 
of the magnetic structure is twice the NaCl-type chemical unit 
cell, so the magnitude of the wave vector of the magnetic modulation 
is given by $k = \frac{1}{2}$ reciprocal lattice units 
(rlu). The primary magnetic reflections are then of the form \peakk\, 
where the $\langle\ldots\rangle$ indicates a permutation over indexes. 
These reflections, which are the \emph{only} ones observed by neutron diffraction, 
are also imaged by resonant x-ray scattering (RXS), via 
the $F^{(1)}$ electric dipole (E1) scattering amplitude
\cite{Longfield01a}. In  
addition, the RXS cross section exhibits F$^{(2)}$ dipole amplitudes 
\cite{Hannon88,Hill96b} which give rise to peaks at positions of the form
\peakkk. The symmetries of the F$^{(1)}$ 
and F$^{(2)}$ terms may be exploited to distinguish respectively between 
single-$\vec{k}$ and $2\vec{k}$ or $3\vec{k}$ structures\cite{Longfield02}. 

Under the constraints of the 
geometric structure factor, the $F^{(2)}$ amplitude projects, 
on a given scattering centre, a \textit{pair} of the \peakk\ 
order parameters which, given inter-site phase coherence yields Bragg
diffraction peaks of the form \peakkk. The respective $F^{(1)}$, $F^{(2)}$
assignments have  
been experimentally verified through the polarisation and azimuthal 
dependence of the scattered photons \cite{Longfield02}. Moreover, even though 
both amplitudes are electric dipole in origin, the contributions from 
the $F^{(2)}$ uranium scattering amplitudes have different matrix 
elements and are distinguished by their resonant energy and lineshape 
from the $F^{(1)}$ profiles \cite{Longfield02,Paixao02,Lovesey03}.

In the course of these experiments an additional group of reflections, 
much weaker than the other two sets described above, of the generic 
form \peakkkk\ have been observed. In this paper we give 
details characterising these reflections and our difficulty in 
explaining them within conventional scattering theory.

\section{EXPERIMENTAL DETAILS AND RESULTS}

Experiments were performed with $\sigma$ incident 
polarisation of the photon beam at the ID20 beamline\cite{ID20web}, ESRF, Grenoble, 
France in the configuration used in previous work\cite{Longfield02}. The studies were carried out on a single crystal of UAs$_{0.8}$Se$_{0.2}$ 
which, above a tetragonal distortion at $T^* \sim 50$~K, exhibits 
the cubic rock salt structure\cite{Longfield01a}. On warming, UAs$_{0.8}$Se$_{0.2}$ is 
known to pass from a magnetic configuration of commensurate ($k=\frac{1}{2}$) to 
incommensurate ($k=0.475)$ wave vector at $T_{0}$ $\sim119$~K, and to the 
paramagnetic state at $T_{N}$ $\sim124$~K \cite{Kuznietz87,Longfield01a}. It has been 
shown, using a combination of both neutron and x-ray techniques 
\cite{Kuznietz87,Longfield02}, to adopt a multi-$\vec{k}$ structure for $T < 124$~K.

\begin{figure}[p!]
\begin{center}
\includegraphics[height=0.6\textheight]{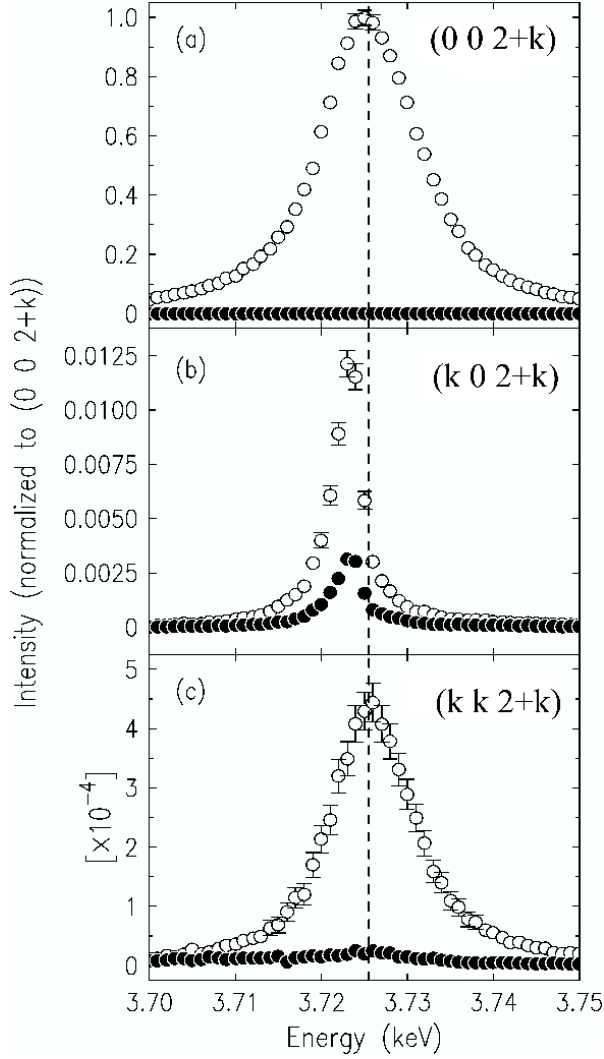}
\caption{
Reflections measured in a single crystal of UAs$_{0.8}$Se$_{0.2}$ with 
photons of energy near the U $M_4$ resonance, which is marked 
with a dashed vertical line. The temperature was 70~K for all 
panels and the wave vector is $k = \frac{1}{2}$. The incident polarization 
is $\sigma$, and using a Au(111) analyser the open (closed) 
points correspond to intensity in the $\pi$ ($\sigma$) channel.
(a) The \peakk\ reflection (0 0 5/2) as observed 
in neutron diffraction. 
Note that it occurs only in the rotated \sigmapi\ channel 
(b) The reflection (1/2 0 5/2) of the form \peakkk\ 
is discussed in detail in Ref. \onlinecite{Longfield02}. It arises because of the 
intrinsic non-collinearity of the $2\vec{k}$ magnetic structure. 
Note that in RXS contributions occur in both polarization channels, 
the energy maximum is shifted and the peaks are narrow (c) A 
new type of reflection (1/2 1/2 5/2) of the form \peakkkk\ 
as discussed in this paper. Intensity is only in the \sigmapi\ channel. 
The energy position and width are similar to case (a) above.}
\label{fig:all-res}
\end{center}
\end{figure}

Representative reflections for the \peakk, \peakkk\ and \peakkkk\ peaks are
illustrated in Figure~\ref{fig:all-res}, where the dependence on incident
photon energy of the scattered 
intensity at the positions (0 0 5/2), (1/2 0 5/2) and (1/2 1/2 
5/2), for both the \sigmasigma\ and \sigmapi\ channels of the cross section 
at T = 70 K are shown. In all cases these energy scans are characterised 
by sharp wave vector profiles indicative of long-range order. 
As Figure~\ref{fig:all-res} shows, the (1/2 1/2 5/2) peak appears \textit{only} in the 
\sigmapi\ channel with a resonant energy and width comparable to 
that of the (0 0 5/2), evidence which already suggests that 
the \peakkkk\ peak may arise from the $F^{(1)}$ dipole (E1) 
amplitude.

\begin{figure}\begin{center}
\includegraphics[width=0.8\columnwidth]{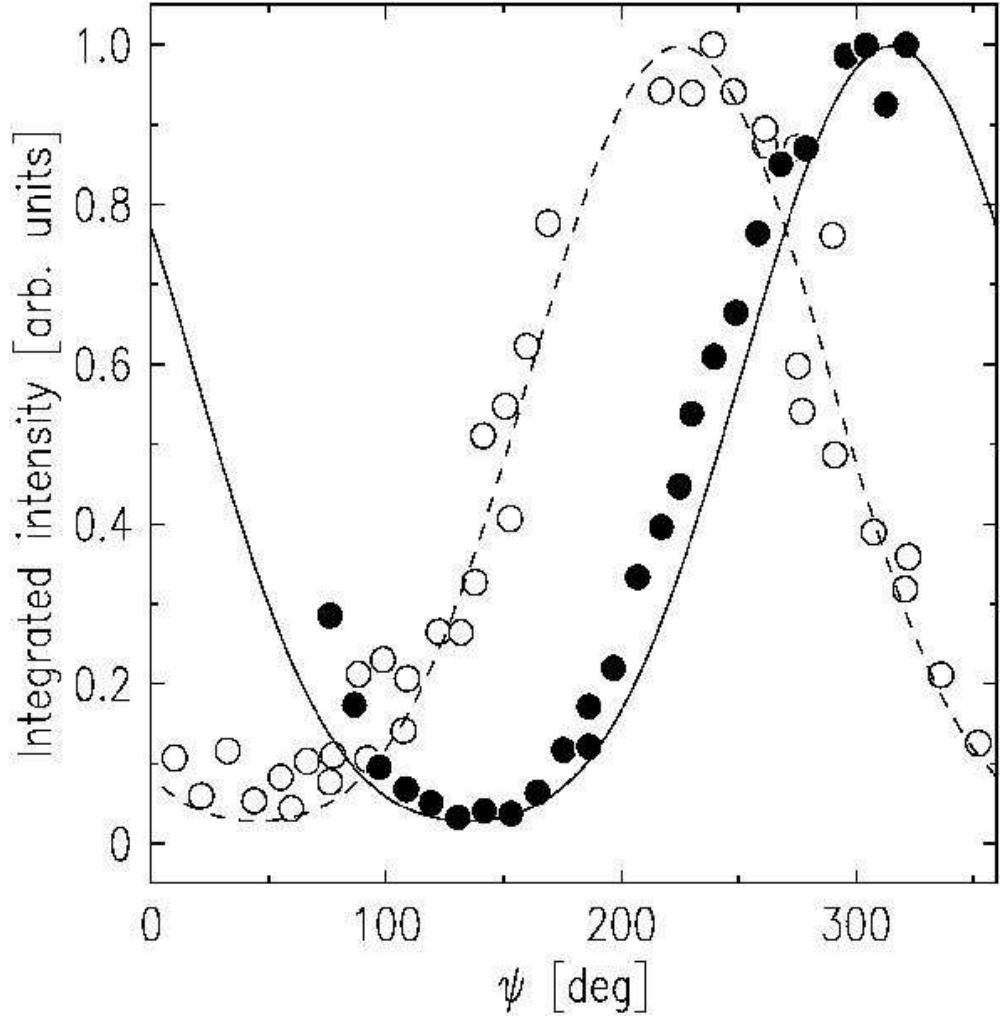}
\caption{
Azimuthal scans in the \sigmapi\ channel bout the scattering vector for the (1/2 1/2 
5/2) (open points) and (--1/2 1/2 5/2) (closed points) reflections. 
The lines correspond to the analysis described in the text. Note 
that for all azimuthal angles the intensity in the scattered \sigmasigma\ channel is zero.
}
\label{fig:kkk-azimuth}
\end{center}\end{figure}

Figure~\ref{fig:kkk-azimuth} shows the azimuthal dependence of the intensity of 
the \peakkkk\ reflections (1/2 1/2 5/2) and (--1/2 1/2 
5/2) in the \sigmapi\ channel. The smooth variation of the intensity 
eliminates multiple scattering as a possible source of these 
peaks. The lines are calculated from the $F^{(1)}$ term of the E1 
cross section, assuming a symmetry breaking vector 
along $\langle111\rangle$, parallel to \peakkkk\, \{i.e. 
along [1 1 1] for the (1/2 1/2 5/2) and along [-1 1 1] for the 
(-1/2 1/2 5/2)\}. The agreement between the data and
this model with only \textit{one} overall scale factor is excellent. These
aspects are discussed further below.

\begin{figure}\begin{center}
\includegraphics[width=0.8\columnwidth]{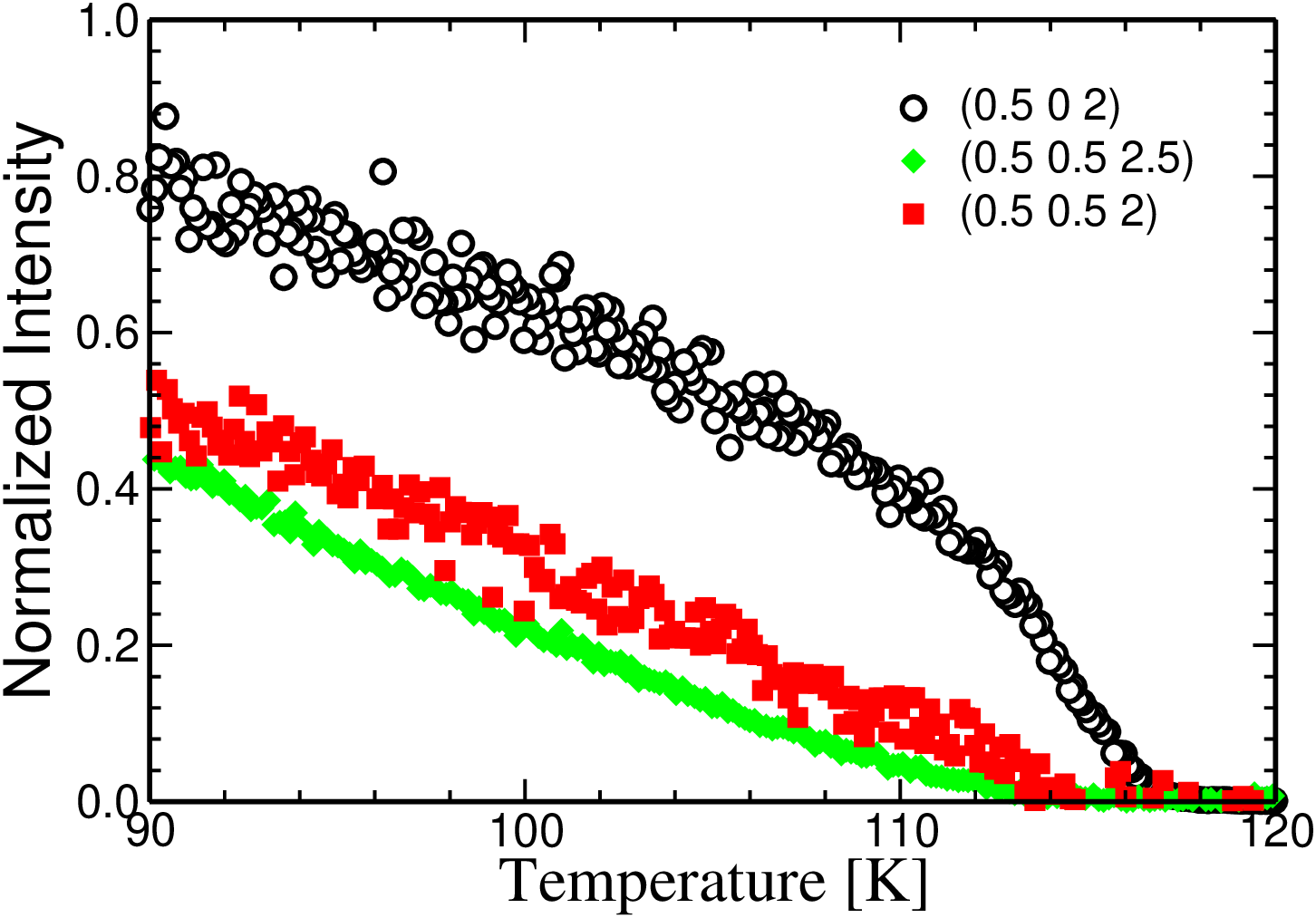}
\caption{Peak intensity of the \peakk\, \peakkk\ and \peakkkk\ (circles
 , squares and diamonds respectively) as a function of temperature.}
\label{fig:kkk-inttemp}
\end{center}\end{figure}

 The temperature dependencies 
of the (0 0 5/2), (1/2 0 5/2) and (1/2 1/2 5/2) peaks are given 
in the Figure~\ref{fig:kkk-inttemp}. The (0 0 5/2) reflection represents 
one primary order parameter whilst the $2\vec{k}$ (1/2 0 5/2) reflection
involves two simultaneously present at each scattering centre and
propagating  
with fixed phase relationship. The temperature dependencies of $I_{\peakk}$,
$I_{\peakkk}$ and $I_{\peakkkk}$ for 70~K$< T <$~117~K in
Figure~\ref{fig:all-res} lie approximately in the ratio $T^{0.4} : T :
T^{1.5}$. 

\begin{figure}\begin{center}
\includegraphics[width=0.8\columnwidth]{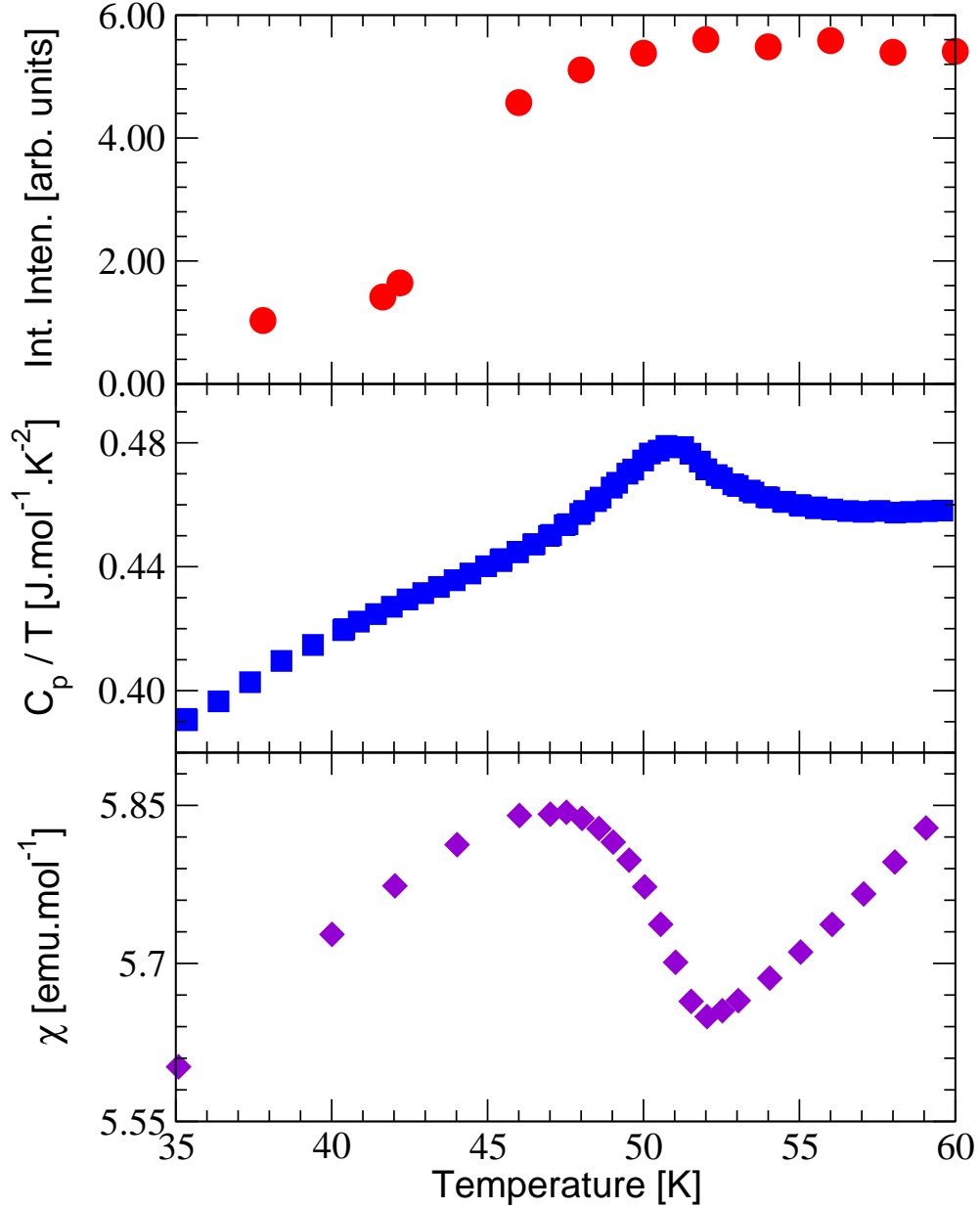}
\caption{Integrated intensity of the $(\frac{1}{2}\frac{1}{2}\frac{5}{2})$
  reflection in the region 35~K$ < T < 60$~K (top panel),
  heat capacity at zero applied magnetic field
  (middle panel) and magnetic suseptibitity in a applied magnetic field
  of 0.1~T (bottom panel).}
\label{fig:cp-and-chi}
\end{center}\end{figure}

Detailed specific heat and magnetisation measurements were made 
on the sample used in the diffraction experiments at 
the User Facility, Institute for Transuranium Elements, Karlsruhe 
\cite{ITUweb}. We show in Figure~\ref{fig:cp-and-chi} the results of (a) the
integrated intensity of 
the (1/2 1/2 5/2) peak, (b) heat capacity at zero field, and 
(c) the susceptibility in a magnetic field of 0.1~T, as a function 
of temperature. The loss of intensity observed in the \peakkkk\ reflection 
is coupled to the nominal $2\vec{k} \Leftrightarrow 3\vec{k}$ transitions seen in both
specific heat  
and susceptibility at $T^*$ (Ref. \onlinecite{Longfield01a}).

\section{DISCUSSION}

Before turning to a possible explanation of this effect, we recapitulate 
the phase diagram of the UAs$_{0.8}$Se$_{0.2}$ as suggested 
by earlier neutron \cite{Kuznietz87} and RXS experiments \cite{Longfield01a}. First, the disappearing 
intensity of the \peakkkk\ peak and the presence of a 
lattice distortion below $T^*\sim 50$~K \cite{Longfield01a}, show that the 
low-temperature state is most likely a 2$\vec{k}$ phase\cite{Kuznietz87}. 
Above $T^*$, high-resolution x-ray 
experiments have not been able to detect any distortion from 
cubic symmetry, this suggests, in agreement with the neutron 
results, that the sample is not in a simple $2\vec{k}$ (or single-$\vec{k}$) 
phase. Previous authors have suggested this is be a $3\vec{k}$
state\cite{Kuznietz87}.

We now examine possible origins of the \peakkkk\ peaks. A simple explanation
would be that at the level of $10^{-4}$ of  
the total volume there are regions that exhibit an ordering at 
a single-$\vec{k}$ wave vector
$\langle\frac{1}{2}\frac{1}{2}\frac{1}{2}\rangle$. This would explain 
the observed energy and azimuthal dependence of the scattering, 
Figures~\ref{fig:all-res} and \ref{fig:kkk-azimuth}, respectively. However,
there are at least two observations 
which contradict such a scenario. First, the similar, 
sharp, $\vec{q}$ widths of \peakk\, \peakkk\ and 
\peakkkk\ reflections are indicative that the \peakkkk\ peaks represent 
(bulk) long-range order. 
Second, the  simple relation of their temperature dependencies to 
the \peakk\ and \peakkk\ peaks for $T^* < T < T_{o}$ 
would then have to be completely fortuitous, which is hard to 
accept. Moreover, single-$\vec{k}$ $(\frac{1}{2}, \frac{1}{2}, \frac{1}{2})$
ordering has, to date, never been reported  
in NaCl-structure uranium compounds. These observations all suggest 
that the \peakkkk\ reflections are intimately related 
to the primary long-range order parameters of the material.

As already noted, both the electric dipole crosssection and the geometric
structure factor of the 
magnetic moment (axial vector) for $\vec{Q} = \peakkkk$ vanish. 
The lowest combination of order parameters with finite geometrical structure
factors is of rank 3, i.e. of the form J$_{x}$J$_{y}$J$_{z}$. For example, a
symmetrised, octupolar operator couples directly to the $\mathrm{F}^{(3)}$ term
of the E2 cross section \cite{Hill96b}, as observed e.g.{\
}in $\mathrm{V_2O_3}$ \cite{Paolasini99,Lovesey00}.  To date however, there is
no evidence of any E2 resonances 
at the actinide $M_{4,5}$ edges, since these would couple to the $g$ states,
with a correspondingly small matrix element. Furthermore, the E2 cross section
would give rise to scattering in both the $\sigma\rightarrow\sigma$
and $\sigma\rightarrow\pi$ polarization channels, and one would expect
the maximum of the resonance to be shifted towards the pre-edge
region, as observed in transition metal \cite{Paolasini99} and rare
earth systems \cite{Gibbs91,Hirota00}. Rather, the energy and polarization
profiles link these reflections to the $\mathrm{F}^{(1)}$ term of the E1 cross
section.

The effective moment direction along $\langle111\rangle$, as indicated by
analysis of the azimuthal dependencies shown in Figure~\ref{fig:kkk-azimuth},
indicates the origin of this resonance lies in the atomic matrix elements. Any
combined lattice distortion or change density wave (CDW) at \peakkk\ with a
magnetic dipole \peakk\ construction is not supported by our
observations. Furthermore, there is no experimental evidence for either a
distortion or CDW in the cubic phase for such hypothetical
constructions\cite{Dmitrienko00}. 

A mechanism to couple an E1 resonance to the octupolar
moment of the valence shell was recently suggested by
\citeauthor{Lovesey03} \cite{Lovesey03}. They found that the rank-2
tensor (E1-$\mathrm{F}^{(2)}$ term in the cross section) observed in
$\mathrm{NpO_2}$ may exist even in the absence of a quadrupolar moment
on the $\mathrm{Np}$ ion. This tensor is constructed from a magnetic octupole
(rank 3) and an induced Zeeman 
splitting in the $3d$ core shell (rank 1). The product of these
tensors contains one of rank 2 which may have been observed in the x-ray
experiment\cite{Paixao02}. 

In a similar framework coupling a rank $(R)$
and rank $(R+1)$ tensor will yield a vector (rank 1) in its product yielding a
\peakkkk\ scattering amplitude. Hypothetical examples include combining a
\peakk\ magnetic moment in the $5f$ valence level with a \peakkk\ quadrupolar
splitting of the $3d$ core levels or vice versa. A still more complex scenario
would be a $J_x$~$J_y$~$J_z$ octupolar moment in the 5f shell (rank 3) with a
rank 2 quadrupolar splitting of the 3d core levels. What physical field would
give the required core level anisotropy is unknown and such high level
mechanisms are currently without foundations. 

We conclude with a brief discussion of the phase diagram. Due to the absence
of any measurable tetragonal 
distortion it was previously suggested that the phase between $T^*$ and $T_O$
is a $3\vec{k}$ state. However, there is no direct evidence for this
assumption. In fact unpublished 
field-dependent specific heat measurements, as well as other 
results are rather difficult to interpret on this basis. Thus, 
the nature of the transition at $T^*$ remains unclear. We suggest the low-temperature (tetragonal) 
configuration to be a state of $2\vec{k}$ domains composed of phase 
coherent pairs of the primary order parameters, in agreement 
with earlier work\cite{Kuznietz87,Longfield01a}. However, as the temperature is raised, the 
tetragonal $2\vec{k}$ phase melts with fluctuations of increasing 
frequency between the possible domains giving, above $T^*$, a dynamic 
state which maintains a cubic environment within which both $2\vec{k}$ 
and $3\vec{k}$ correlations of the primary order parameters may coexist. 
At present we have no direct evidence on the lifetime of either 
$2\vec{k}$ or $3\vec{k}$ correlations above $T^*$. The rapid nature of the 
RXS technique (temporal resolution of $\sim 10^{-15}$ s) may 
be of importance if the coherence time scales are on the scale 
of the inverse bandwidth. 

Whilst this paper eliminates many of the most obvious explanations 
for the presence of the \peakkkk\ reflections clearly 
more work is needed to explain their observation at both \emph{qualitative}
and \textit{quantitative} levels. 
We hope the observations and discussion will stimulate further 
experiments and theoretical studies of such multi-$\vec{k}$ systems. 

\begin{acknowledgments}

We thank Ted Forgan, Christian Vettier, Anne Stunault, and Fran\c{c}ois 
de Bergevin for interesting discussions and Matt Longfield for 
his help in the early part of these investigations. SBW, PJ and EB would like
to thank the European Commission for support in the frame of the `Training and
Mobility of Researchers' programme. 

\end{acknowledgments}

\bibliography{res-uasse-kkk}

\end{document}